\documentclass[a4paper,useAMS,usenatbib,onecolumn,fleqn]{mn2e}

\usepackage{graphics}
\usepackage[dvipdfm]{graphicx}
\usepackage{amsmath,amssymb,bm,color}

\usepackage{tabularx}
\usepackage{epstopdf}
\usepackage{fancyvrb}
\usepackage{longtable}
\usepackage{arydshln}

\def\be{\begin{equation}}
\def\ee{\end{equation}}
\def\ba{\begin{eqnarray}}
\def\ea{\end{eqnarray}}
\def\bv{\begin{verbatim}}
\def\ev{\end{verbatim}}

\def\dTb{\delta T_{\rm b}}
\def\Tb{T_{\rm b}}
\def\Ts{T_{\rm s}}
\def\Tk{T_{\rm k}}

\def\Ta{T_{\alpha}}

\def\Tcmb{T_{\rm CMB}}

\def\yc{y_{\rm c}}
\def\ya{y_{\alpha}}

\def\Lya{{\rm Ly-}\alpha}

\def\kB{k_{\rm B}}
\def\hp{h_{\rm p}}
\def\A10{A_{10}}
\def\v10{\nu_{10}}
\def\k10{\kappa_{10}}
\def\dTb{\delta T_{\rm b}}

\def\He3   {^{3}{\rm He}}
\def\He3I  {^{3}{\rm He}}
\def\He3II {^{3}{\rm He}}
\def\He3III{^{3}{\rm He}}
\def\iHe3{^{3}{\rm He}^{+}}

\def\nsiso{n_{s}^{\rm iso}}

\def\rcdm{r_{\rm cdm}}
\def\rbar{r_{\rm bar}}

\def\omc{\Omega_{\rm c}}
\def\omb{\Omega_{\rm b}}
\def\oml{\Omega_{\Lambda}}

\def\As{A_{\rm s}}
\def\ns{n_{\rm s}}
\def\as{\alpha_{\rm s}}

\def\mK{{\rm mK}}
\def\m{{\rm m}}

\def\hr{{\rm hr}}

\def\MHz{{\rm MHz}}

\def\deg{{\rm deg}}
\def\arcmin{{\rm arcmin}}
\def\deg2{{{\rm deg}^{2}}}
\def\arcmin2{{{\rm arcmin}^{2}}}

\def\mH{{m_{\rm H}}}

\def\fsky{f_{\rm sky}}

\def\zmax{z_{\rm max}}

\def\Mmax{M_{\rm max}}
\def\Mmin{M_{\rm min}}

\def\Tk{T_{\rm k}}
\def\Tb{T_{\rm b}}
\def\dTb{{\delta T_{\rm b}}}

\title[]{ 
  Constraining isocurvature perturbations with the 21cm emission from
  minihaloes}
\author[]{
Yoshitaka Takeuchi$^{1}$\thanks{E-mail:yoshitaka@nagoya-u.jp} 
and 
Sirichai Chongchitnan$^{2}$ 
\vspace{5pt} \\
$^{1}$Department of physics, Nagoya University, 
Naogya 464-8602, Japan \\
$^{2}$Department of Physics and Mathematics, University of Hull, Cottingham Road, Hull, HU6 7RX, United Kingdom \\
}

\begin{document}

\date{Accepted, Received; in original form }

\pagerange{\pageref{firstpage}--\pageref{lastpage}} \pubyear{2014}

\maketitle

\label{firstpage}

\begin{abstract}

We investigate the effects of isocurvature perturbations on the 21cm radiation from minihaloes (MHs) at high redshifts and examine constraints on the isocurvature amplitude and power spectrum using the next generation of radio telescopes such as the Square Kilometre Array. We find that there is a realistic prospect of observing the isocurvature imprints in the 21cm emission from MHs, but only if the isocurvature spectral index is close to $3$ (\textit{i.e.} the spectrum is blue). When the isocurvature fraction increases beyond $\sim 10\%$ of the adiabatic component, we observe an unexpected decline in the 21cm fluctuations from small-mass MHs, which can be explained by the incorporation small MHs into larger haloes. We perform a detailed Fisher-matrix analysis, and conclude that the combination of future CMB and 21cm experiments (such as CMBPol and the Fast-Fourier-Transform Telescope) is ideal in constraining the isocurvature parameters, but will stop short of distinguishing between CDM and baryon types of isocurvature perturbations, unless the isocurvature fraction is large and the spectrum is blue.

\end{abstract}


\section{Introduction}
\label{sec:intro}

Recent measurements of the anisotropies in the cosmic microwave background (CMB) by the \textit{Planck} satellite have placed constraints of unprecedented accuracy on the amplitude of the primordial density fluctuations \citep{Planck-I:2013,Planck-XXII:2013}. Planck also revealed that these fluctuations are consistent with having originated from  \textit{adiabatic} initial conditions, characterized by the constancy of the ratios of density contrasts of various particle species in the early Universe (see \cite{Kodama:1984, Bardeen:1980} for reviews). This is in agreement with previous CMB measurements by the \textit{WMAP} satellite \citep{WMAP9a,WMAP9b}. On the other hand, if the aforementioned ratios of density contrasts are \textit{not} constant, the fluctuations are said to be generated from  \textit{isocurvature} initial conditions, of which there are four types, namely, the cold-dark-matter (CDM), baryon, neutrino-density and neutrino-velocity isocurvature perturbations \citep{Bucher:2000}. Constraints from Planck limit any isocurvature contributions to the CMB temperature anisotropies to less than $\sim10$ percent.

The simplest model of inflation involving a single, slowly rolling scalar field predicts that density fluctuations are generated from purely adiabatic initial conditions. Hence, the detection of any isocurvature contribution would be a window to novel physical mechanisms in the inflationary era. Such mechanisms include the curvaton mechanism \citep{Lazarides:2004,Langlois:2004,Moroi:2005a,Moroi:2005b,Ichikawa:2008a,Langlois:2008}, 
the axion and gravitino CDM \citep{Rajagopal:1991,Covi:2001,Covi:2002,Brandenburg:2004} 
and 
the modulated reheating scenarios \citep{Dvali:2004,Kofman:2003,Ichikawa:2008b,Takahashi:2009a,Takahashi:2009b} 
as well as various combinations of such scenarios. In most of these models,  a large isocurvature fraction can be produced at the expense of the introduction of a few additional parameters \citep{Moroi:2002,Lyth:2003a,Lyth:2003b,Beltran:2008,Moroi:2009,Takahashi:2009a}.

According to our current understanding of cosmology, inflation-stretched primordial quantum fluctuations subsequently grow via gravitational instability into the observed cosmic structures. One of the earliest cosmic structures to form were minihaloes (MHs), which are virialized haloes of dark and baryonic matter with typical mass $10^4-10^8$ $M_\odot$, and temperature $\lesssim 10^4$ K, at very high redshift ($z\sim 6-20$). Minihaloes typically host a high density of neutral hydrogen, which can be detected by the 21cm absorption/emission line due to the transition of the hydrogen atom from a parallel to anti-parallel spin state. MHs are typically at such high temperatures that their 21cm signal appears in emission with respect to the CMB \citep{Iliev:2002}.  The 21cm signals from MHs give us information on the small-scale density fluctuations at high redshifts, and their detection will therefore lead to a deeper understanding of small-scale physics during the earliest structure-formation epoch.

The 21cm signal from MHs has previously been studied by \cite{Chongchitnan:2012}, who showed that the 21cm emissions from MHs are a sensitive probe of primordial non-Gaussianity, due to a strong dependence of the MH number density and bias on the amplitude of non-Gaussianity.  \cite{Tashiro:2013} calculated the 21cm fluctuations due to MHs in cosmic wakes produced by cosmic strings.

In this paper, we present a new probe of isocurvature fluctuations using the 21cm signal from MHs. We will show that the fluctuations in the 21cm emission from MHs are a viable probe of isocurvature fluctuations. We also give forecasts on the isocurvature fraction and spectral index using the next generation of large arrays of radio interferometers, which are expected to measure the cosmic 21cm signals over a wide range of redshifts, from the cosmic Dark Ages ($z\sim30-50$) down to the Epoch of Reionization (EoR) at $z\sim6$. Such radio surveys include: the Low-Frequency Array (LOFAR\footnote{http://www.lofar.org}), the Murchion Widefield Array (MWA\footnote{http://www.mwatelescope.org}), and the Giant Metrewave Radio Telescope (GMRT\footnote{http://gmrt.ncra.tifr.res.in}), all of which focus on $6 \la z \la 30$, as well as more ambitious future arrays such as the Square Kilometre Array (SKA\footnote{http://www.skatelescope.org}), 
and the Fast Fourier Transform Telescope (FFTT) \citep{FFTT}, which can probe  the radio Universe at $z \ga 30$.

There have only been a handful of works exploring the link between 21cm cosmology and isocurvature perturbations:  \cite{Barkana:2005} and \cite{Lewis:2007} discussed the prospects for differentiating between the CDM and baryon isocurvature fluctuations using 21cm signals. Further work by \cite{Kawasaki:2011} showed that 21cm surveys can effectively probe the difference between CDM and baryon isocurvature fluctuations if the spectrum of isocurvature perturbations is strongly blue tilted (we revisit this claim later). \cite{Gordon:2009}  investigated the constraints on isocurvature modes from 21cm observations, focusing on the so-called compensated isocurvature perturbations.

This paper is organized as follows: we summarize the 21cm radiation from minihaloes and its sensitivity to the presence of isocurvature modes in 
Sec.~\ref{sec:21cm}. The effects of isocurvature modes on the \textit{fluctuations} of this signal are explained in Sec.~\ref{sec:iso}. 
Forecasts on the constraints of isocurvature parameters from future radio surveys are discussed in
Sec.~\ref{sec:forecast} and \ref{sec:result}.  
Finally, Sec.~\ref{sec:discuss} and \ref{sec:conc} contain further discussions and a summary of our main conclusions.

Throughout this work, we assume a flat Universe and adopt the cosmological parameters from Planck \citep{Planck-XVI:2013}.

\section{21cm emission from minihaloes}
\label{sec:21cm}

The 21cm spectral line can appear in either emission or absorption against the CMB  depending on the spin temperature, $T_s$, determined by the balance between collisional and radiative excitations of the hydrogen atoms.  The interactions between a hydrogen atom and photons, electrons and other atoms  couple the spin temperature to the temperatures of the surrounding gas and radiation field as \citep{Field:1958} 
\begin{equation}
  \Ts = \frac{\Tcmb + \ya \Ta + \yc \Tk}{1 + \ya + \yc} ~ ,
\label{eq:tspin}
\end{equation}
where $\Ta$ is the colour temperature of Ly$\alpha$ photons, $\Tk$ is the kinetic temperature, and $\ya$ and $\yc$ are the radiative and collisional excitation efficiencies \citep{Madau:1997}. We  assume that bright UV and X-ray sources have yet to form or that the MHs are isolated from such sources. Thus, we can neglect the radiative coupling and set $y_\alpha=0$.

The amplitude of the 21cm signal from a virialized halo depends on the density profile, velocity and temperature of the halo. We adopt as our model the \textit{truncated isothermal sphere} (TIS) \citep{Shapiro:1999,Iliev:2001}, in which a minihalo of a given mass is described by its radius $r_t$, temperature $\Tk$, 
density profile $\rho(r)$ and velocity dispersion $\sigma_V$. 
In this model, each minihalo is modelled as a non-singular sphere of dark matter and baryons in virial and hydrostatic equilibrium, so that $\rho(r)$ describes both the dark matter and gas profiles.

The observed brightness temperature along a line of sight, through a halo at comoving distance $r$ from the center of the halo, is given by 
\begin{equation}
  \Tb(r) = \Tcmb(z) e^{-\tau(r)} + \int_0^{\tau(r)} \Ts e^{-\tau'}d\tau' ~ ,
\end{equation} 
where $\tau(r)$, the total optical depth of neutral hydrogen to photons at frequency $\nu$, can be expressed as \citep{Furlanetto:2002} 
\begin{equation}
  \tau(\nu) = \frac{3 c^2 A_{10} T_{*}}{32 \pi \nu_0^2} 
  \int_{-\infty}^{\infty} 
  \frac{n_{\rm HI}(\ell) \phi(\nu,\ell)}{\Ts(\ell)} dR. 
\end{equation}
Here, $R$ and $\ell$ are radial comoving distances satisfying $\ell^2 = R^2 + (\alpha r_t)^2$; $\alpha$ is the impact parameter in unit of $r_t$, and $n_{\rm HI}$ is the number density of neutral hydrogen. 
$\phi(\nu)$ is the intrinsic Doppler-broadened line profile  given by
\begin{equation}
  \phi(\nu) = \frac{1}{\Delta \nu \sqrt{\pi}} \exp \left[ - \left(\frac{\nu-\nu_0}{\Delta \nu}\right)^2 \right] ~ ,
\end{equation}
with $\Delta \nu = (\nu_0/c) \sqrt{2\kB\Tk/\mH}$.

When the line profile is unbroadened; $\phi(\nu) = \delta(\nu - \nu_0)$, the optical depth corresponds to that of the IGM at redshift $z$ and can be expressed as \citep{Madau:1997}
\begin{equation}
  \tau_{\rm IGM}(\nu; z) = \frac{3 c^3 A_{10} T_{*} n_{\rm HI}(z)}{32 \pi \nu_0^3 \Ts(z) H(z)} ~ ,
\end{equation}
where $A_{10}$ and $\nu_{10}$ are the spontaneous decay rate and the rest-frame frequency for the 21cm transition, $T_{*}$ is the equivalent temperature defined as $T_{*} \equiv \hp \nu_{10} / \kB$. 
The total optical depth can be written as 
\begin{equation}
  \tau(\nu, R) = \tau_{\rm IGM}(\nu) + \frac{3 c^2 A_{10} T_{*}}{32 \pi \nu_0^2} 
  \int_{-\infty}^{R} 
  \frac{n_{\rm HI}(\ell') \phi(\nu,\ell')}{\Ts(\ell')} dR' ~.  
\end{equation}
The first and second terms represent the contributions from IGM and the MH respectively. 
\\

The differential 21cm brightness temperature, $\delta T_{\rm b}$, measured with respect to the CMB temperature, is given by 
\begin{equation}
  \dTb = \frac{1}{1+z} \left( \frac{\int dA \Tb(r)}{A} - \Tcmb(z) \right) ~ , 
\end{equation}
where $\Tb$ is averaged over the halo cross-section $A = \pi r_t^2$. 
The mean 21cm emission from an ensemble of MHs in the mass range [$\Mmin$,$\Mmax$] is thus given by \citep{Iliev:2002}
\begin{equation}
  \overline{\dTb} = \frac{c (1+z)^4}{\nu_0 H(z)}
  \int_{M_{\rm min}}^{M_{\rm max}} \Delta \nu_{\rm eff} \dTb(M) A 
  \frac{dn}{dM} dM ~ ,
\label{eq:bar_dTb}
\end{equation}
where $\nu_{\rm eff} = [\phi(\nu_0)(1+z)]^{-1}$ is the effective redshifted line width. We take $M_{\rm max}$ to be the virial temperature corresponding to temperature $10^4$ K, and $M_{\rm min}$ to be the Jeans mass, $M_{\rm J}$.

The \textit{rms} fluctuations in the 21cm emission for a pencil-beam survey with bandwidth $\Delta \nu$ and angular size $\Delta \theta$ is given by 
\begin{equation}
  \langle \delta T_{\rm b}^2 \rangle^{1/2} = \sigma_{\rm p}(z,\Delta \nu,\Delta \theta) \beta(z) \overline{\dTb}(z) ~ ,
\end{equation}
where $\sigma_{\rm p}$ is the variance in a cylinder and $\beta$ is the flux-weighted average of the halo bias.

The variance in a cylinder is given by
\begin{equation}
  \sigma_{\rm p}(z, \Delta \nu, \Delta \theta) 
  = 
  2 \pi
  \int \frac{dk_z}{k_z} \left[ \frac{k_z^3 P(k_z)}{2\pi^2} \right] 
  \int_{1/R}^{\infty} {dk_r} 
  \left[ \frac{2}{k_r R(z)} 
    j_0 \left( \frac{k_z L(z)}{2} \right) 
    J_1 \left(k_r R(z) \right) \right]^2 \, ,
\end{equation}
where $L$ and $R$ represent respectively the width along the line of sight and the spatial resolution of survey, $P(k)$ is the matter power spectrum, and $\nu_0=1.42$ GHz is the rest-frame frequency for a 21cm transition.

The flux-weighted average of the halo bias is given by
\begin{equation}
  \beta(z) = \frac{\int_{M_{\rm min}}^{M_{\rm max}} b(M,z) {\cal F}(M) \frac{dn}{dM} dM}
       {\int_{M_{\rm min}}^{M_{\rm max}} {\cal F}(M) \frac{dn}{dM} dM} ~ ,
\label{eq:beta}
\end{equation}
where ${\cal F}(M) \propto \Tb r_t^2 \sigma_{\rm V}$ is the effective flux from the MHs and $b(M,z)$ is the halo bias. We adopt the bias expression of \cite{Sheth:2001} in this work. According to \cite{Iliev:2001}, the non-linear bias approach of \cite{Scannapieco:2002} can be robustly reproduced by the linear bias obtained in \cite{Mo:1996}. We have checked that our choice of $b(M,z)$ closely reproduces the result using the bias of \cite{Mo:1996} or that of \cite{Chongchitnan:2012a} in the Gaussian case.

\section{Effects of isocurvature perturbations}
\label{sec:iso}

\begin{figure}
\begin{center}
\includegraphics[clip,keepaspectratio=true,width=0.85
  \textwidth]{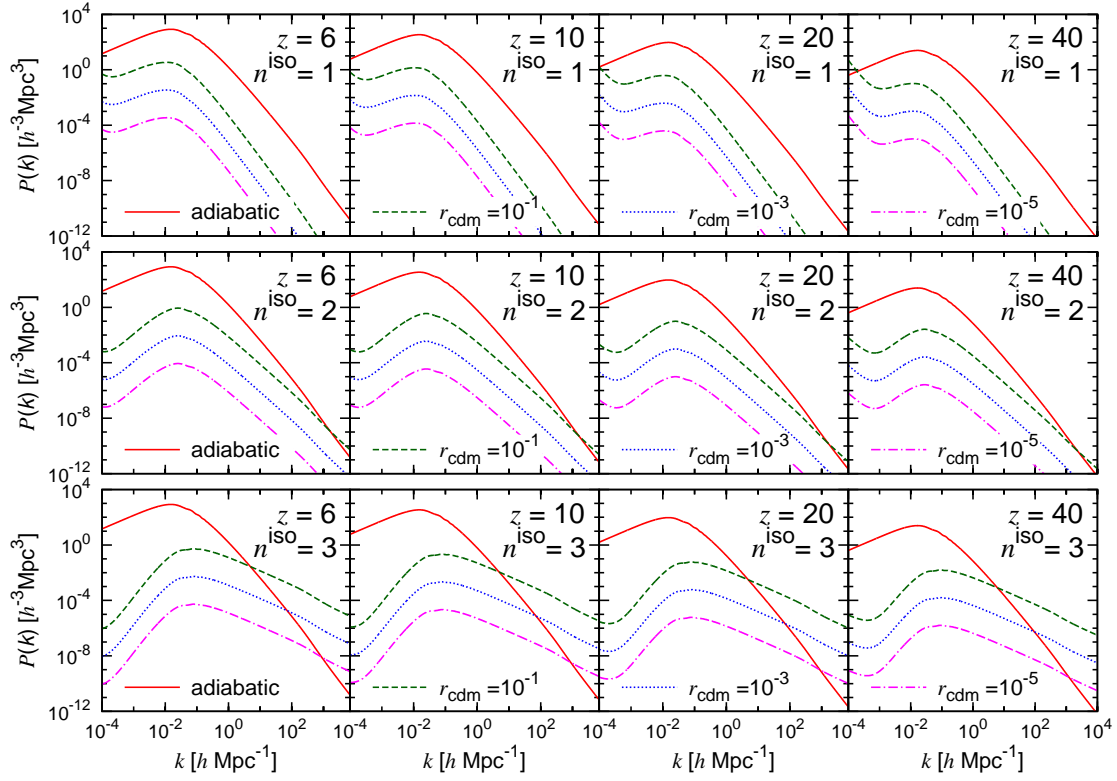}
\end{center}
\caption{The matter power spectra generated by adiabatic or pure CDM isocurvature fluctuations, at redshifts $z=6$, 10, 20 and 40 (from left column to right). The spectral indices of the isocurvature mode are  as $\nsiso=1$, 2 and 3 (from top row to bottom). In each panel, the different curves represent the matter power spectrum of the adiabatic fluctuations (solid/red) and the CDM isocurvature fluctuations with $\rcdm=10^{-1}$ (dashed/green), $10^{-3}$ (dotted/blue) and $10^{-5}$ (dot-dashed/magenta). The isocurvature spectra shown have no contribution from adiabatic fluctuations. 
}
\label{fig:pks}
\end{figure}

\begin{figure}
\begin{center}
\includegraphics[clip,keepaspectratio=true,width=0.85
  \textwidth]{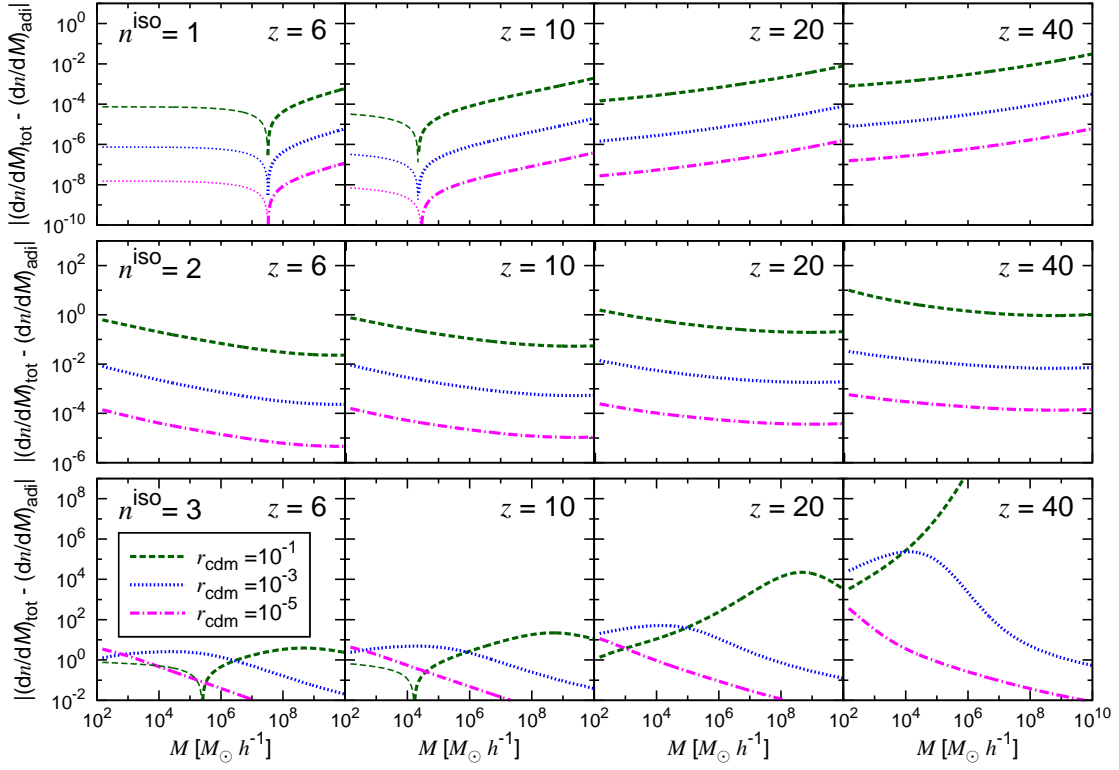}
\end{center}
\caption{Deviation of the halo mass function for the case with CDM isocurvature fluctuations from the pure adiabatic case at redshifts $z=6$, 10, 20 and 40 (from left to right). $(dn/dM)_{\rm tot}$ represents the halo mass function for the total (adiabatic+CDM isocurvature) fluctuation; $(dn/dM)_{\rm adi}$ is the mass function for the pure adiabatic case. Thick (thin) lines represent positive (negative) values. The spectral indices of the isocurvature perturbations are  $\nsiso=1$, 2 and 3, (from top to bottom). Different line types represent different values of $r_{\rm cdm}$ (same as those in Fig. \ref{fig:pks}). 
}
\label{fig:dndm}
\end{figure}

As a first step in calculating the effects of isocurvature perturbations, we parametrize the primordial power spectrum for isocurvature fluctuations as 
\begin{equation}
  {\cal P}_{S_i}(k) \equiv {\cal P}_{S_i}(k_0) \left( \frac{k}{k_0} \right)^{n_s^{i}-1} ~ , 
\end{equation}
where $i={c}$ or ${b}$, indicating CDM and baryon isocurvature modes. ${\cal P}_{S_i}(k_0)$ and $n_s^{(i)}$ are, respectively, the amplitude and the spectral index for the mode $i$ defined at reference scale $k_0$. In this paper, we take $k_0=0.05$ Mpc$^{-1}$.

We define the primordial isocurvature fractions as 
\begin{equation}
  r_{\rm cdm} \equiv \frac{{\cal P}_{S_{c}}(k_0)}{{\cal P}_{\zeta}(k_0)} ~ ,
  \hspace{12pt}
  r_{\rm bar} \equiv \frac{{\cal P}_{S_{b}}(k_0)}{{\cal P}_{\zeta}(k_0)} ~ ,
\end{equation}
where ${\cal P}_{\zeta}(k_0)$ is the amplitude of the primordial power spectrum for the adiabatic (curvature) perturbation $\zeta$. 
For simplicity, we adopt a single value for both the spectral indices of CDM and baryon isocurvature spectra, and denote this by $\nsiso$ (\textit{i.e.} $n_{s}^{c} = n_{s}^{b} = \nsiso$). 
The total matter isocurvature perturbation, $S_{\rm m}$ is given by the combination of the isocurvature fluctuations (with respect to radiation) in CDM and in baryon ($S_{\rm c}$ and $S_{\rm b}$), as $S_{\rm m}=f_{\rm c}S_{\rm c} + f_{\rm b} S_{\rm b}$, where $f_{\rm c}=\Omega_{\rm c}/\Omega_{\rm m}$ and $f_{\rm b}=\Omega_{\rm b}/\Omega_{\rm m}$. It is worth noting that if  CDM and baryons contribute equally to the total isocurvature fluctuations, the initial amplitude of power spectrum for the baryon mode must be larger than that of CDM by a factor of $(\Omega_{\rm c}/\Omega_{\rm b})^2$.

The evolution of isocurvature fluctuations is influenced by two main factors; evolution of the metric perturbations and the amplitudes of initial fluctuations. Although the evolution of the metric perturbations is almost same between the CDM and baryon isocurvature modes, the difference in the initial fluctuations between the CDM and baryon isocurvature modes can lead to observable effects, as will be shown in this work (see also \cite{Kawasaki:2011}).

Let us first consider the effects of isocurvature modes on the matter power spectra. The case of a pure CDM isocurvature mode is shown in Figure~\ref{fig:pks} at redshifts $z=6$, 10, 20 and 40 (from left column to right) with  isocurvature fraction $r_{\rm cdm}=10^{-1}$ (dashed/green), $10^{-3}$ (dotted/blue) and $10^{-5}$ (dot-dashed/magenta), with varying spectral index $\nsiso=1$, 2 and 3 (from top row to bottom). The solid/red line in each panel shows the adiabatic spectrum.

We see that for very blue-tilted spectrum ($\nsiso=3$), the effects of the CDM isocurvature mode can be identified clearly on small-scales, whereas the contribution from a scale-invariant isocurvature spectrum ($\nsiso=1$) is much smaller than the adiabatic component even with relatively large isocurvature fractions.
\\

Figure ~\ref{fig:dndm} shows the changes in the halo mass function due to the contribution from isocurvature fluctuations with respect to the adiabatic case. Each curve is derived from the corresponding matter power spectrum shown in Figure~\ref{fig:pks}, using the prescription of \cite{Press:1974}. We clearly see that the effects of the isocurvature modes are prominent on small mass scales and at high redshifts. 
In particular, we see that very blue-tilted isocurvature spectra ($n_s^{\rm iso}=3$) exhibit very different features from the other spectra with $n_s^{\rm iso}=1$ or 2. The changes in the halo mass function do not vary monotonically with increasing fractional amplitudes $r_{\rm cdm}$. In general, blue-tilted isocurvature spectra show enhanced fluctuations on small-scales, and lead to the increase of the number of small haloes. However, if the contribution of isocurvature modes increases beyond some critical value, small haloes can become incorporated into larger haloes. This explains the unexpected features seen in the last row of Figure ~\ref{fig:dndm}, where the abundances of massive haloes are enhanced, but those of   smaller-mass haloes are suppressed. Such a feature appears in the typical mass range of MHs, \textit{i.e.} $[\Mmin,\Mmax]$, and it is expected that the 21cm signal from MHs will also exhibit such a trend. 
\\

\begin{figure}
\begin{center}
\includegraphics[clip,keepaspectratio=true,width=0.90
   \textwidth]{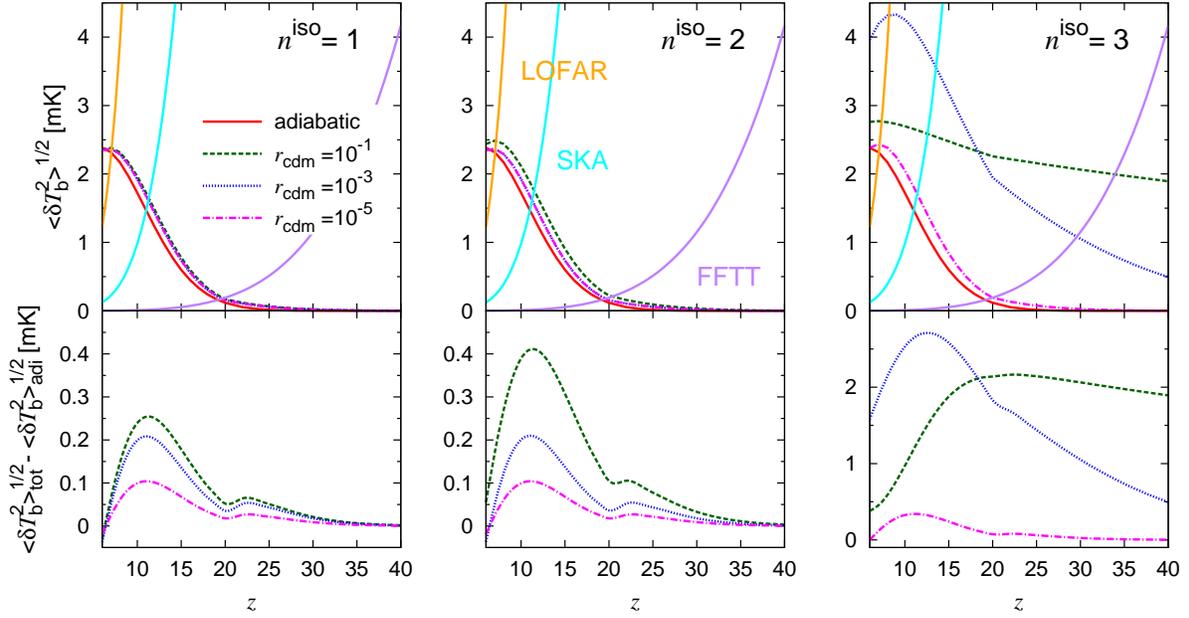}
\end{center}
\caption{\textit{Top panels}: The \textit{rms} fluctuations in the 21cm emission from MHs, $\langle \dTb^{2} \rangle^{1/2}$, with the sensitivity curves for LOFAR (orange), SKA (cyan) and FFTT (purple). The curves in each panel represent the same cases as in the previous figures.  \textit{Lower  panels}: Deviation from the adiabatic case, with isocurvature spectral indices $\nsiso=1$, 2 and 3, (from left to right). 
  $\langle \dTb^{2} \rangle^{1/2}_{\rm tot}$ and $\langle \dTb^{2} \rangle^{1/2}_{\rm adi}$ respectively represent the signal for the case with adiabatic and CDM isocurvature fluctuations and that for the case with adiabatic fluctuations alone. 
  The unusual trend in the last column is discussed in the text.
}
\label{fig:dTb21cm}
\end{figure}

Finally, we calculate the \textit{rms} fluctuations in the 21cm emission from MHs, $\langle \delta T_{\rm b}^2 \rangle^{1/2}$, as a function of redshift  (Figure~\ref{fig:dTb21cm}). Again, we assume the contribution from only the CDM type of isocurvature fluctuations. The sensitivity curves are for LOFAR, 
SKA and FFTT (details of the sensitivities are explained in the next section).  
We see that when the isocurvature spectrum is flat ($n_s^{\rm iso}$=1), the difference in $\langle \delta T_{\rm b}^2 \rangle$ compared to the adiabatic mode is $\la 10^{-4}$ mK even with $r_{\rm cdm}=0.1$. This suggests that such isocurvature components would be extremely difficult to observe through MHs. Even with  bluer isocurvature spectrum ($n_s^{\rm iso}=2$), the difference is still small: the model with $r_{\rm cdm}=0.1$ enhances the signal by not much more than a few percent around the $z\sim10$. 

If the isocurvature spectrum is very blue ($n_s^{\rm iso}=3$),  large differences can be seen, especially at high redshifts. However, a slight trend reversal is seen around $z \la 20$, where $r_{\rm cdm}=10^{-3}$ boosts the signal more effectively than when $r_{\rm cdm}=10^{-1}$. This again can be understood in terms of the incorporation of small-mass MHs into larger haloes, as previously discussed.

Our calculation shows that the detection of isocurvature contribution to the fluctuations in the 21cm MH emission is possible with future telescopes such as the SKA and FFTT. If isocurvature fluctuations have a very blue spectrum with $\rcdm \simeq 10^{-3}$, such isocurvature signals may be detected at low redshifts even by LOFAR. However, further increase in $\rcdm$ suppresses the  signal at $z\la20$ due to the incorporation of small MHs into larger haloes. We shall discuss other uncertainties in the calculation of the 21cm signal from MHs in Sec.~\ref{sec:discuss}.

\section{Forecasts}
\label{sec:forecast}

We now perform a Fisher-matrix analysis on the cosmological parameters derived from measurements of the CMB and the fluctuations in the 21cm signal  from MHs. 
We define the total Fisher matrix by combining the CMB and the 21cm surveys as 
\begin{equation}
  F_{\alpha \beta} = F_{\alpha \beta}^{({\rm CMB})} + F_{\alpha \beta}^{({\rm 21cm)}} ~ , 
\end{equation}
where $\alpha$, $\beta$ refer to the model parameters, and $F_{\alpha \beta}^{({\rm CMB})}$ and $F_{\alpha \beta}^{({\rm 21cm})}$  represent the contributions from the CMB and 21cm observations. 
We adopt following 12 parameters in our analysis;
\begin{equation}
  {\bm p} = \{    
  \omb h^2,\ \omc h^2,\ \oml,\ \tau^{\rm reion},\ \ns,\ \As, 
  w,\ Y_p,\ \as,\ 
  \rcdm, \rbar,\ \nsiso
  \} ~ , 
\label{eq:params}
\end{equation}
where 
$\omb$, $\omc$ and $\oml$ are the density parameters for baryons,
CDM and cosmological constant respectively;
$h$ is the dimensionless Hubble constant; 
$w$ is the equation of state for dark energy; 
$Y_p$ is the primordial abundance of Helium; 
$\tau^{\rm reion}$ is the optical depth at the EoR\footnote{We treat $\tau^{\rm reion}$ as a model parameter only in the CMB measurement since $\tau^{\rm reion}$ does
not affect the 21cm signals in our analysis.} 
; 
$\ns$ and $\As$ are the spectral index and the amplitude of the primordial
power spectrum for the adiabatic mode;
$\as$ is the running of the spectral index; 
$\rcdm$ and $\rbar$ are CDM and baryon isocurvature fractions;
$\nsiso$ is the spectral index for the isocurvature perturbations.

\subsection{CMB Fisher matrix}

The Fisher matrix for a CMB survey is given by \citep{Tegmark:1997}
\begin{equation}
  F_{\alpha \beta}^{({\rm CMB})} = f_{\rm sky}^{\rm CMB} \sum_{\ell=2}^{l_{\rm max}} \frac{2\ell + 1}{2} {\rm Tr} 
  \left[ {\bf C}_{\ell;\alpha} {\bf C}_\ell^{-1} {\bf C}_{\ell;\beta} {\bf C}_\ell^{-1} \right] ~ ,
\end{equation}
where $f_{\rm sky}^{\rm CMB}$ is the sky coverage of the CMB survey, ${\bf C}_\ell$ is the covariance matrix and ${\bf C}_{\ell;\alpha}$ represents its derivative with respect  parameter $p_\alpha$; ${\bf C}_{\ell;\alpha} \equiv \partial {\bf C}_{\ell}/\partial p_\alpha$. The CMB observables include the temperature anisotropies ($T$), the E-mode polarization $(E)$, and the CMB lensing potential ($\psi$). The covariance matrix constructed from these observables is given by
\begin{equation}
  {\bf C}_\ell \equiv \left(
  \begin{array}{lll}
    C_\ell^{TT} + N_\ell^{TT} & C_\ell^{TE} & C_\ell^{T\psi} \\ 
    C_\ell^{TE} & C_\ell^{EE} + N_\ell^{EE} & C_\ell^{E\psi} \\ 
    C_\ell^{T\psi} & C_\ell^{E\psi} & C_\ell^{\psi\psi} + N_\ell^{\psi\psi} 
  \end{array}
  \right) ~ ,
\end{equation}
where $C_\ell$ and $N_\ell$ represent the angular power spectrum and the noise spectrum respectively. For simplicity, we assume that the cross-correlation between the E-mode polarization and the CMB lensing potential can be neglected (\textit{i.e.} $C_\ell^{E\psi}=0$)\footnote{This is because E-mode polarization is generated via Thomson scattering around the last-scattering surface, whereas the sources of CMB lensing are the large-scale structures between us and the last scattering surface. However, such correlation, though small, is not exactly zero since the E-mode polarization can also be generated during the EoR, and structures in this era can also act as lensing sources \citep{Lewis:2011}.}.

The noise spectrum for a CMB experiment is given by \citep{Knox:1995}
\begin{equation}
  N_{\ell}^{T,P} = \left[  \sum_\nu
  \left\{ \left( \Delta_{\nu}^{T,P} \theta_{\rm FWHM}\right)^2 
    e^{-\ell(\ell + 1) \theta_{\rm FWHM}^2/8\ln 2} \right\}^{-1} \right]^{-1} ~ ,
\end{equation}
where $\Delta_\nu^{T,P}$ denotes the sensitivity of the temperature or polarization measurement, and $\theta_{\rm FWHM}$ represents the angular resolution (the so-called full-width at half-maximum). We calculate the noise spectrum for the lensing-potential measurement using the formalism outlined in \cite{Hu:2002} and \cite{Okamoto:2003}.   In particular, we assume the projected sensitivities of the \textit{CMBPol} mission \citep{CMBPol}, with $f_{\rm sky}^{\rm CMB}$=1 and $\ell_{\rm max}$=4000.  We use specifications for a mid-cost CMBPol (EPIC-2m type) mission, as shown in Table~\ref{tb:cmb}.

\begin{table}
\begin{center}
\begin{tabular}{cccccc}
\hline \hline \vspace{-8pt} \\
\makebox[2cm]{$\nu$ [MHz]} &
\makebox[2cm]{$\theta_{\rm FWHM}$ [arcmin]} &
\makebox[2cm]{$\Delta_\nu^{T}$ [$\mu$K arcmin]} &
\makebox[2cm]{$\Delta_\nu^{P}$ [$\mu$K arcmin]} \\
\hline

 45  & 17  & 5.85  & 8.27 \\
 70  & 11  & 2.96  & 4.19 \\
 100 & 8   & 2.29  & 3.24 \\
 150 & 5   & 2.21  & 3.13 \\
 220 & 3.5 & 3.39  & 4.79 \\
\hline \hline
\end{tabular}
\end{center}
\caption{The specifications for a mid-cost \textit{CMBPol} (EPIC-2m type) mission adopted in this paper. Here $\nu$ refers to the frequency of each channel, $\theta_{\rm FWHM}$ is the angular resolution, $\Delta_\nu^T$, and $\Delta_\nu^P$ are the sensitivities for the temperature and polarization measurements.}
\label{tb:cmb}
\end{table}

\subsection{21cm Fisher matrix}

For a 21cm survey, we define the Fisher matrix as
\begin{equation}
  F_{\alpha \beta}^{({\rm 21cm})} = f_{\rm sky}^{\rm 21cm} \sum_{i}\sum_{\rm pixel}
  \left( \frac{\partial S^{i}}{\partial p_\alpha} \right)
  \frac{1}{2(S^{i} + N^{i})^2}
  \left( \frac{\partial S^{i}}{\partial p_\beta} \right) 
  ~ ,
\end{equation}
where $i$ runs over all redshift slices, $f_{\rm sky}^{\rm 21cm}$ is the sky coverage for the 21cm survey, $S^{i}$ and $N^{i}$ represent the signal and noise in the $i$-th redshift slice. We define the signal and the noise as $S^{i} \equiv \langle \delta T_{\rm b}^2(z_i) \rangle^{1/2}$ and $N^{i} \equiv \delta T_N(z_i)$ respectively, and $\delta T_N(z)$ is given by \citep{Furlanetto:2006}
\begin{equation}
  \delta T_N(z) = 20 {\mK} 
  \frac{10^4 \m^2}{A_{\rm tot}} \left[ \frac{10'}{\Delta \theta} \right]^2
  \left[ \frac{1+z}{10} \right]^{4.6} 
  \left[ \frac{\MHz}{\Delta \nu} \frac{100 \hr}{t_{\rm int}} \right]^{1/2} \, ,
\end{equation}
where
$A_{\rm tot}$ is the effective collecting area of the radio array, 
$\Delta \theta$ is the angular resolution,
$\Delta \nu$ is the frequency bandwidth,
and 
$t_{\rm int}$ is the total integration time. The sensitivity curves shown in Figure~\ref{fig:dTb21cm} assume $A_{\rm tot}=10^4$ m$^2$ (LOFAR), $A_{\rm tot}=10^5$ m$^2$ (SKA) and $A_{\rm tot}=10^7$ m$^2$ (FFTT), with $t_{\rm int}=1000$ hours in all cases.

As a fiducial survey, we use the specifications of FFTT, with $A_{\rm tot}$=$10^7$ m$^2$, $\Delta \theta$=9 arcmin, $\Delta \nu$=1 MHz, and $t_{\rm int}$=$1000$ hours.

\section{Results}
\label{sec:result}
\subsection{The $r_{\rm cdm}-r_{\rm bar}$ plane}

\begin{figure}
\begin{center}
\includegraphics[clip,keepaspectratio=true,width=0.9
   \textwidth]{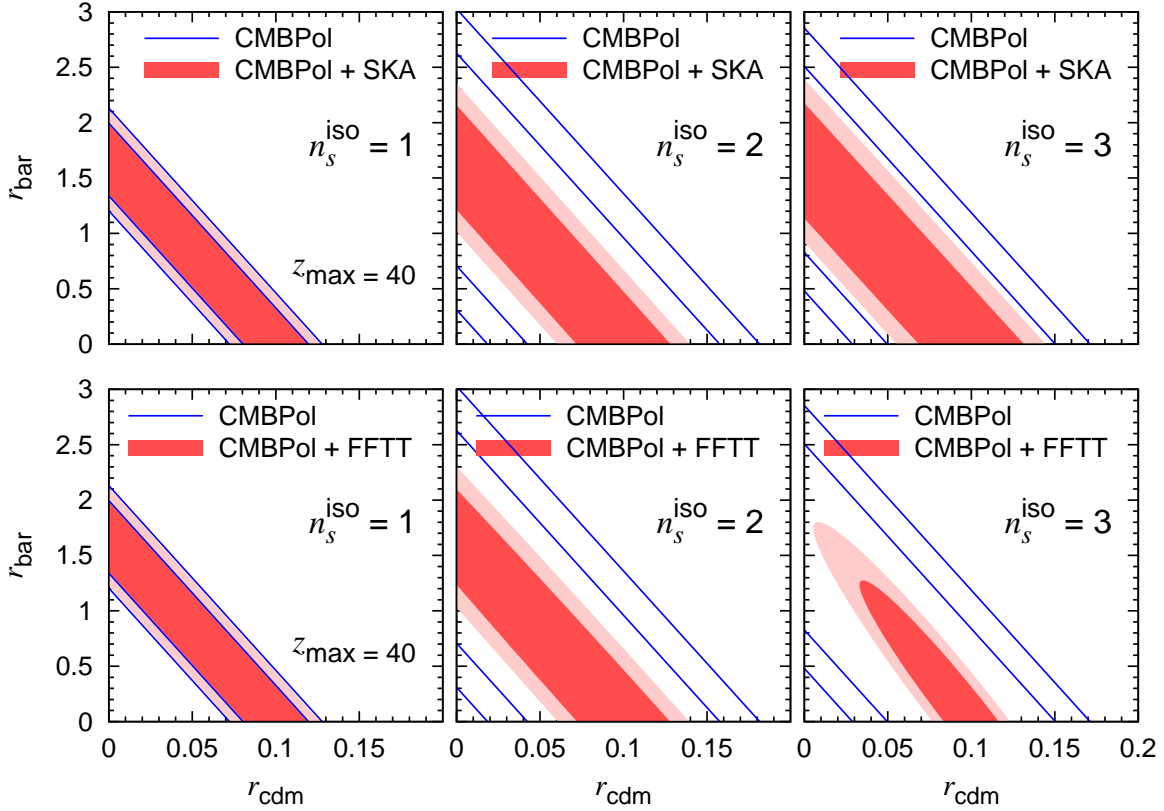}
\end{center}
\caption{Projected 1$\sigma$ (68 \%) and 2$\sigma$ (95 \%) constraints in the $\rcdm-\rbar$ plane from the CMB alone (solid/blue line) and CMB+21cm (shaded/red region). We assume CMBPol specifications and the SKA (top panels) or FFTT (bottom panels) for the observation of 21cm MH signal. For the fiducial model, we used $\nsiso=1$, 2 and 3 (from left to right), and the fiducial isocurvature fractions are ($\rcdm$,\,$\rbar$)=(0.1,\,0) in all cases. 
}
\label{fig:cont_iso}
\end{figure}

Figure~\ref{fig:cont_iso} summarises the results from our Fisher analysis. The contours show the projected 1$\sigma$ (68\%) and 2$\sigma$ (95\%)  constraints in the  $r_{\rm cdm}-r_{\rm bar}$ plane expected from CMBPol alone and from combining with either the  SKA or FFTT. 
We perform the analysis over the redshift range $6\leq z\leq40$ in equally spaced bins centred at $z_i$ with bin separation $\Delta z_i =1$. Within each bin, we assume the bandwidth resolution of $\Delta\nu = 1 $ MHz.

As shown in the previous section, the contribution from a scale-invariant ($n_s^{\rm iso}=1$) isocurvature spectrum to the 21cm MH signal is small. This is also evident from the contours, which are only modestly tightened by when 21cm constraints are added to those from CMBPol. The improvement is more dramatic for bluer isocurvature spectra, especially in the bottom right panel, where we can see that it is possible to break parameter degeneracies by the combining CMB and 21cm constraints. 
Comparing the constrains from SKA and FFTT, both sets of constraints show similar results, except in the case  with $\nsiso=3$, where the constraint from FFTT is clearly much tighter than that from the SKA.

The constraint on the case with $\nsiso=3$ from a combination of CMBPol and FFTT shows larger error of $\rbar$ than that of $\rcdm$. The difference of the amplitude of errors between CDM and baryon isocurvature fluctuations comes from the fact that the baryon isocurvature fluctuations are required the larger amplitude by a factor of $(\Omega_{\rm c}/\Omega_{\rm b})^2$ to realize the same amount of isocurvature fluctuations with that of CDM.

\subsection{Dependence on $z_{\rm max}$}

\begin{figure}
\begin{center}
\includegraphics[clip,keepaspectratio=true,width=0.85
  \textwidth]{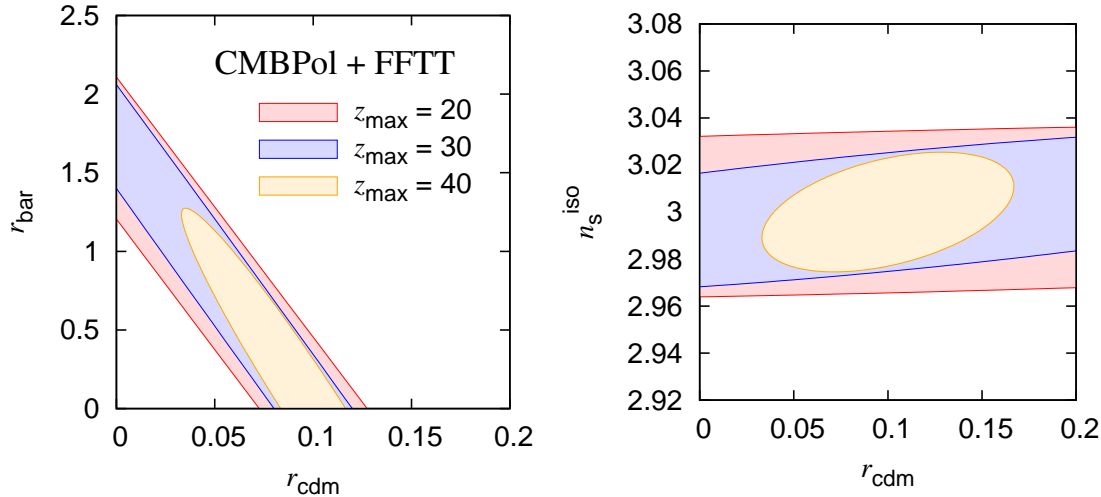}
\end{center}
\caption{The effects of varying $z_{\rm max}$ (maximum redshifts probed by the 21cm experiment) on the $1\sigma$ constraints from the CMBPol+FFTT in the $r_{\rm cdm}-r_{\rm bar}$ plane (\textit{left panel}) and the $r_{\rm cdm}-n_s^{\rm iso}$ plane (\textit{right}). The fiducial model is the adiabatic model plus CDM isocurvature with $\rcdm=0.1$ and $\nsiso=3$. Different colour contours represent different values of $\zmax$, where $\zmax=20$ (outer/red), 30 (middle/blue), 40 (inner/orange).}
\label{fig:cont_zmax}
\end{figure}

Next, we consider the dependence of the constraints on the redshift range used in the Fisher analysis. Figure~\ref{fig:cont_zmax} shows the 1$\sigma$ contours expected from CMBPol+FFTT, where the maximum redshift varies from $\zmax=20$ to 40. We show the results in both the $r_{\rm cdm}-r_{\rm bar}$ and the $r_{\rm cdm}-n_s^{\rm iso}$ planes. The contours suggest that if information up to $\zmax \sim 40$ can be utilized, there is some hope of differentiating the CDM and baryon isocurvature perturbations. Incidentally, we noted that the constraints from the SKA are saturated when $\zmax \sim 20$, beyond which point the signal-to-noise ratio for the SKA falls below one.

\subsection{Information in redshift slices}

\begin{figure}
\begin{center}
\includegraphics[clip,keepaspectratio=true,width=0.9
  \textwidth]{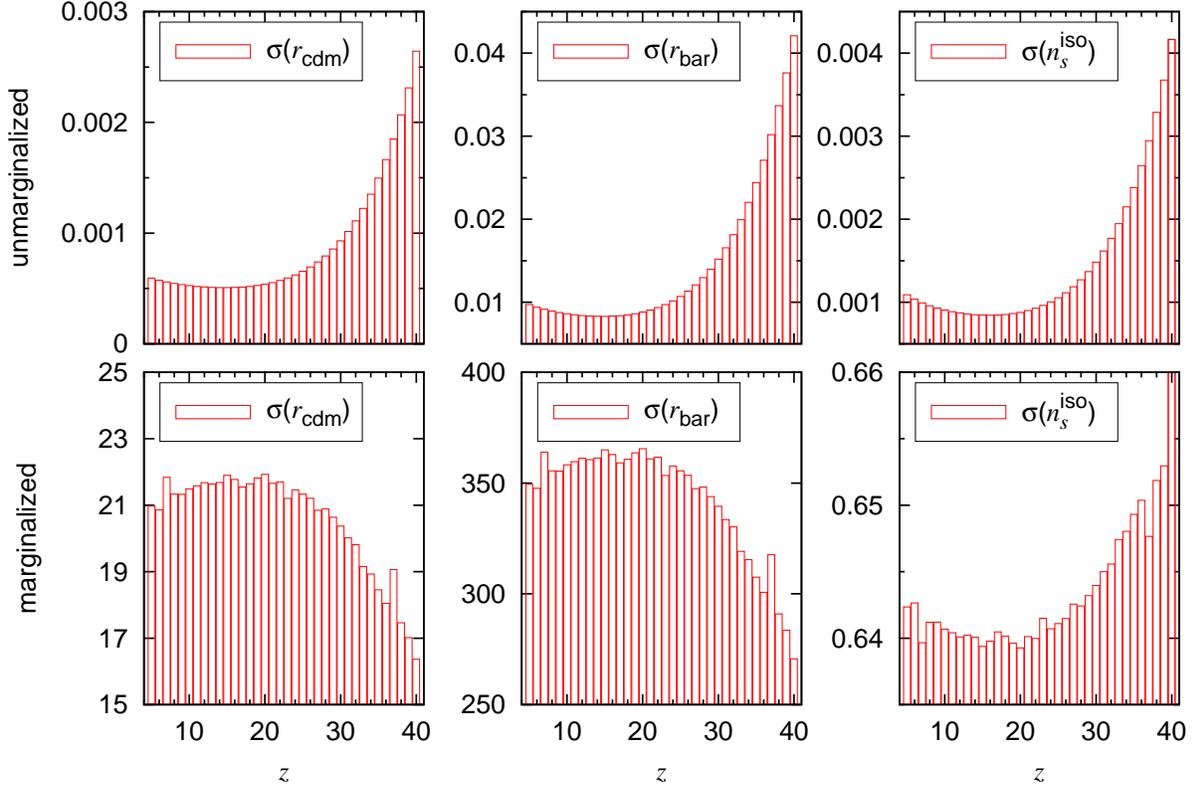}
\end{center}
\caption{The unmarginalized (top panels) and marginalized (bottom panels) 1$\sigma$ errors in each redshift slice. The fiducial model is the adiabatic model plus CDM isocurvature fluctuations with ($\rcdm$,$\rbar$)=(0.1,0) and $\nsiso=3.0$, and specifications of CMBPol+FFTT are assumed. The spatial and frequency resolutions are  $\Delta \theta$=9 arcmin and $\Delta \nu$=1 MHz respectively. The observed trends are discussed in the text.
}
\label{fig:sigma_z}
\end{figure}

We further investigate the information content in each redshift slice, to  determine which redshifts constrain the isocurvature perturbations most effectively. In Figure~\ref{fig:sigma_z},  we plot the diagonal components for the inverse Fisher matrix, $({\bm F}^{-1})_{\alpha\alpha}$, in the cases when the errors are marginalized (top row), and unmarginalized (bottom row), with $\alpha=\rcdm, \rbar$ and $\nsiso$. The 21cm survey is again taken to be the FFTT, and we use the CMBPol prior in each redshift slice.

For the unmarginalized error, the minimum of the 1$\sigma$ errors appears around $z=20$, which is slightly higher than the peaks of the 21cm signal from MHs ($z\sim10$). This is because the effects of isocurvature modes are more prominent at higher redshifts. 
On the other hand, the marginalized errors show the opposite trends from the unmarginalized errors for $\rcdm$ and $\rbar$. This is due to the strong degeneracy between CDM and baryon isocurvature modes, as well as degeneracies with the other cosmological parameters. As discussed in \cite{Kawasaki:2011}, the differences between CDM and baryon isocurvature modes become more distinct on large scales. Since observations at higher redshift include larger correlation lengths with the same angular scale, the marginalized errors in $\rcdm$ and $\rbar$ are reduced with increasing redshift.

\subsection{Dependence on $f_{\rm sky}$}

\begin{figure}
\begin{center}
\includegraphics[clip,keepaspectratio=true,width=0.85
  \textwidth]{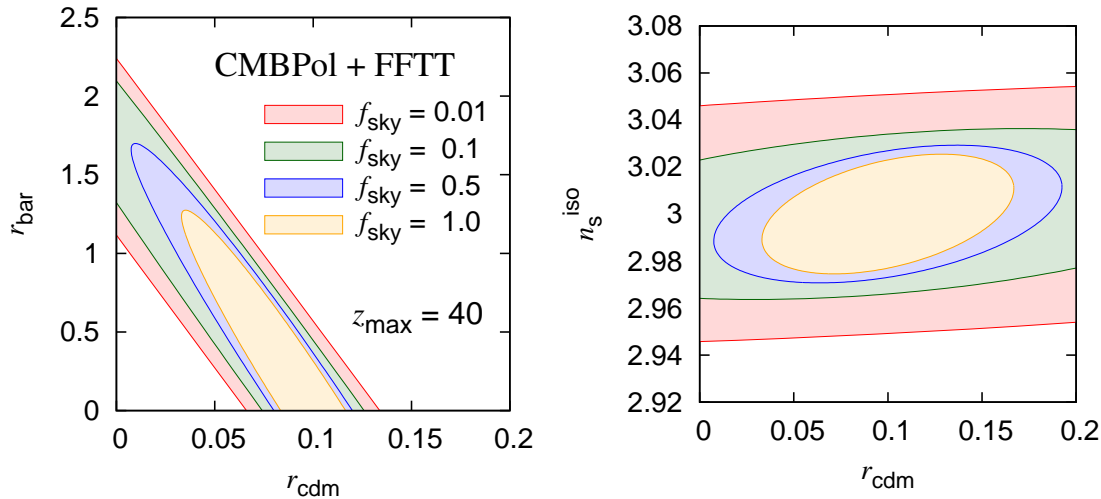}
\end{center}
\caption{The effects of varying $f_{\rm sky}$ on the projected 68 \% CL isocurvature constraints from CMBPol+FFTT, where $\fsky=0.01$ (outermost/red), 0.1 (green), 0.5 (blue) and 1.0 (innermost/orange). The fiducial model is the adiabatic model plus CDM isocurvature fluctuations with $\rcdm=0.1$ and $\nsiso=3$, and  $\zmax=40$.}
\label{fig:cont_area}
\end{figure}

Finally, we examine the dependence of the isocurvature constraints on the sky coverage. We compare the constraints from the fluctuations in the 21cm MH emission using $\fsky=0.01$, 0.1, 0.5 and 1, and show the 1$\sigma$ error contours in Figure~\ref{fig:cont_area}. The constraints in both the $r_{\rm cdm}-r_{\rm bar}$ plane (\textit{left panel}) and the $r_{\rm cdm}-n_s^{\rm iso}$ plane (\textit{right}) are shown.

We see that in order to rule out $\rcdm = 0$ with $68\%$ confidence, more than half the sky must be surveyed using the combined CMBPol and FFTT and probing MHs up to $z_{\rm max}=40$. We also see that the spectral index $n_s^{\rm iso}$ can be constrained with accuracy up to a few percent if $\fsky$ is at least $0.1$.

\section{Discussions}
\label{sec:discuss}

MHs are generally small, nonlinear objects and their dynamics are governed by nonlinearity on small scales. $N$-body simulations are, therefore, the most reliable way to study their dynamics. It is, however, a challenging task to resolve small MHs in $N$-body simulations (see \textit{e.g.} \cite{Shapiro:2004,Richardson:2013} for previous simulations).

Following \citep{Chongchitnan:2012}, we now discuss two additional factors  concerning MH dynamics which may affect the results presented in the previous Section; (i) uncertainty in the halo mass function, and (ii) uncertainty of mass range of MHs. Our results are summarized in Figure \ref{fig:dTb_sys} and \ref{fig:cont_mfs}.

\subsection{Uncertainty in the mass function}

The left column of Figure~\ref{fig:dTb_sys} shows the 21cm fluctuations for $(\rcdm,\rbar)=(0.1,0)$ (top) and $(\rcdm,\rbar)=(10^{-3},0)$ (bottom) using various prescriptions for the halo mass functions, namely, \cite{Press:1974} (PS), \cite{Sheth:1999} (ST), \cite{Tinker:2008} and \cite{Warren:2006}.

We see that when the isocurvature fraction is small ($\rcdm=10^{-3}$), the PS and Tinker prescriptions give similarly high amplitudes of the signal from MHs, whereas the Warren and ST prescriptions both give lower amplitudes. The trends are reversed for high-redshifts. These behaviours agree with those found by \cite{Chongchitnan:2012}. When the isocurvature fraction is large ($\rcdm=0.1$), the PS prescription shows an unexpectedly low 21cm signal at $z\lesssim10$. Nevertheless, these mass functions generally predict similar trends and amplitudes that do not differ significantly.

The left column in Figure~\ref{fig:cont_mfs} shows the $1\sigma$ contours in the $r_{\rm cdm}-r_{\rm bar}$ plane (top) and the $r_{\rm cdm}-n_s^{\rm iso}$ plane (bottom) using CMBPol+FFTT, when different mass functions are adopted.  We observe that mass functions which predict larger  amplitudes of $\langle \dTb^2 \rangle^{1/2}$ show relatively tighter constraints, as one might expect. The PS mass function shows a particularly tight constraint in the $\nsiso-\rcdm$ plane, and one could interpret this as an overestimation of the constraining power of the MHs on isocurvature parameters when the PS formalism is used.

\subsection{Uncertainty in $\Mmin$}

Another uncertainty is the mass range $[M_{\rm min}, M_{\rm max}]$ of MHs. Whilst we have so far taken $M_{\rm min}$ to be the Jeans mass, $M_J$, large relative velocities between dark matter and baryons can cause the advection of baryons out of dark matter potential and  result in $\Mmin > M_{\rm J}$ \citep{Tseliakhovich:2011,McQuinn:2012}. The uncertainty in $M_{\rm max}$ is, in comparison, far less serious since the sharp decline in the halo mass function ensures that the number of very massive MHs is suppressed.

\begin{figure}
\begin{center}
\includegraphics[clip,keepaspectratio=true,width=0.7
  \textwidth]{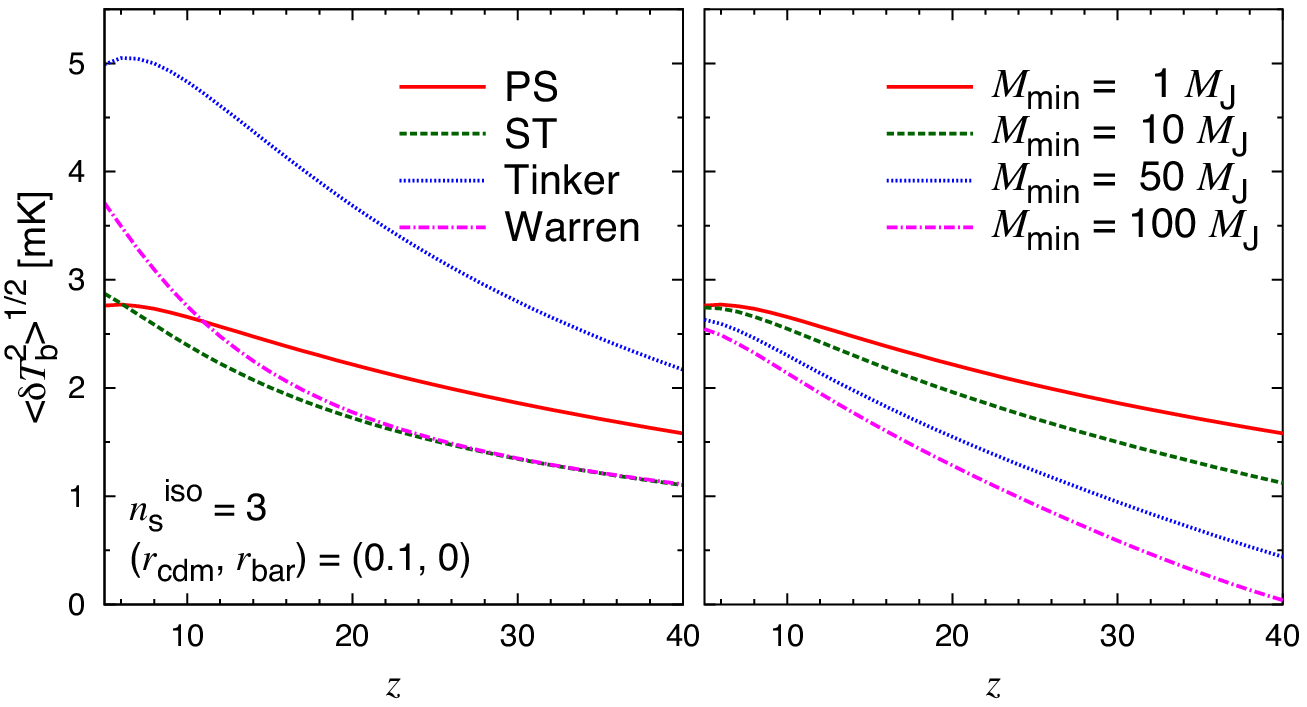}
\includegraphics[clip,keepaspectratio=true,width=0.7
  \textwidth]{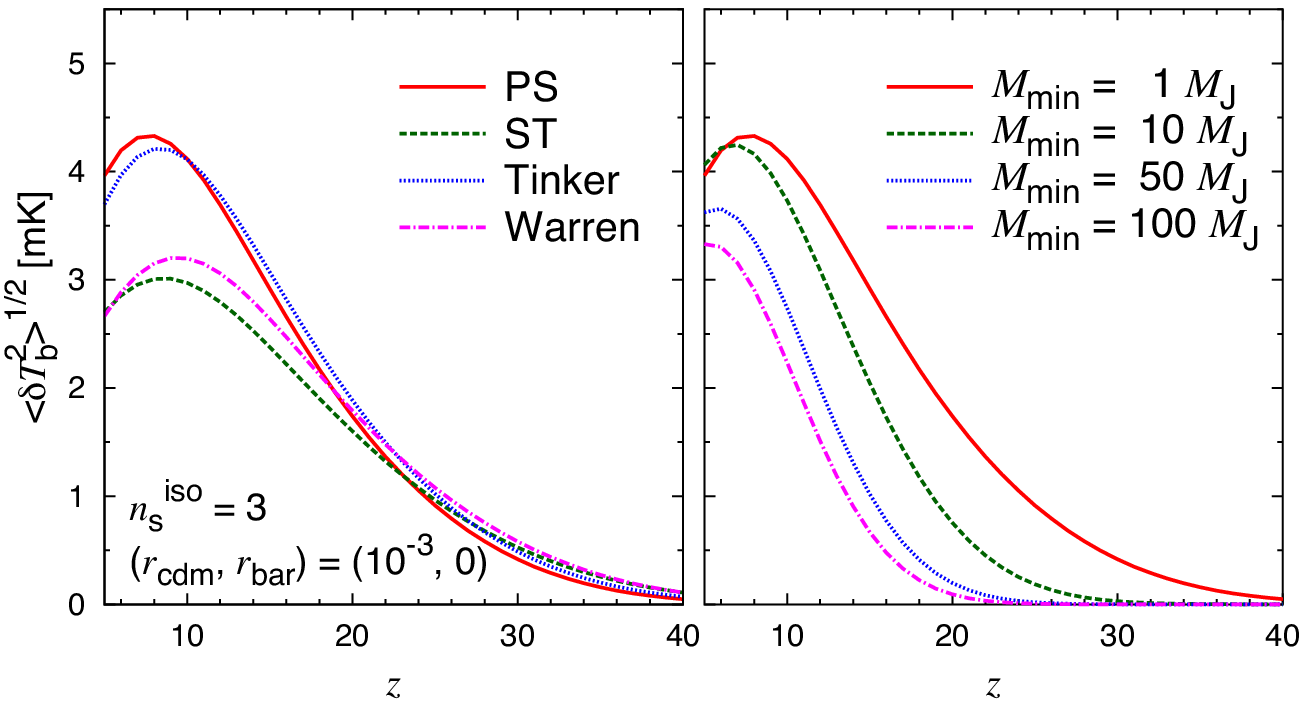}
\end{center}
\caption{The effects of changing the halo mass function (left) or the minimum mass of MHs, $\Mmin$  (right), on the fluctuations in the 21cm emission from MHs. Top panels show the cases with large CDM isocurvature fraction ($\rcdm=0.1$) whilst the lower panels show the cases with $\rcdm=10^{-3}$.
}
\label{fig:dTb_sys}
\end{figure}

\begin{figure}
\begin{center}
\includegraphics[clip,keepaspectratio=true,width=0.75
  \textwidth]{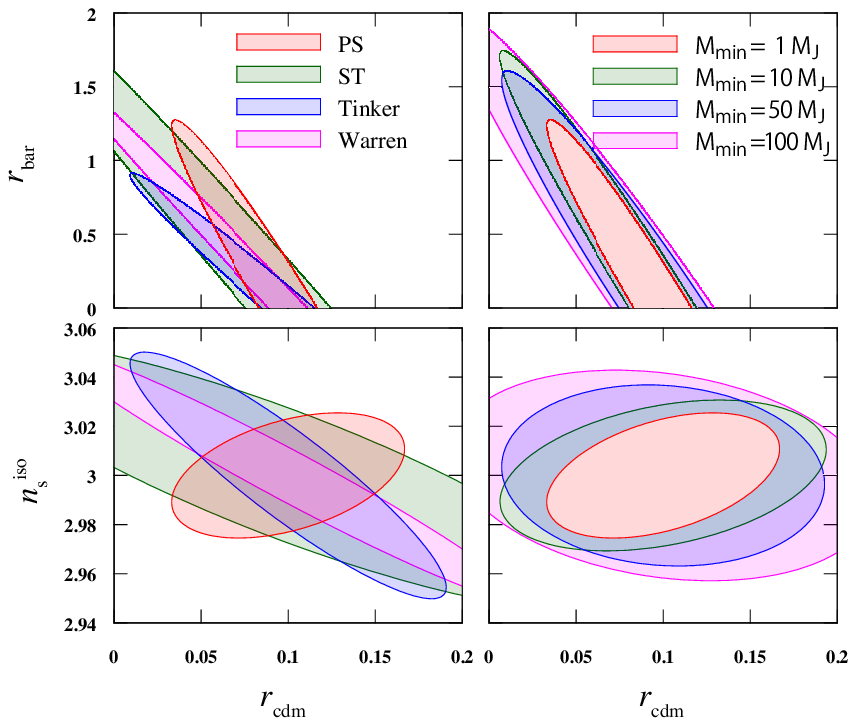}
\end{center}
\caption{ The effects of changing the the halo mass function (left) or the minimum mass of MHs, $\Mmin$  (right), on the projected $1\sigma$ constraints in the $r_{\rm cdm}-r_{\rm bar}$ plane (top) and the $r_{\rm cdm}-n_s^{\rm iso}$ plane (bottom).}
\label{fig:cont_sys}
\label{fig:cont_mfs}
\label{fig:cont_Mmin}
\label{fig:cont_Lya}
\end{figure}

The 21cm signal from MHs using $M_{\rm min}=$ 10, 50 and 100 times the Jeans mass are shown in the right column of Figure~\ref{fig:dTb_sys}. We see that the increasing $\Mmin$ suppresses the signal over all redshifts, with the suppression more prominent at higher redshifts. This is, of course, due to the reduction in the number of MHs. The right column in Figure~\ref{fig:cont_sys} shows the corresponding effects on the $1\sigma$ constraints, which, as expected, become poorer when $\Mmin$ is increased. \\

There are of course other uncertainties in the theoretical modelling of MHs which have not been pursued here, including the interaction of MHs with external UV sources through $\Lya$ pumping \citep{Chongchitnan:2012}, as well as deviations of MHs from the TIS profile. Indeed, dark matter and gas in MHs could possibly take on different, more complex profiles than those postulated by the TIS model. A more numerical approach than presented here would be required however (see \textit{e.g.} \cite{Ricotti:2009,Ricotti:2007}).

\section{Conclusions}
\label{sec:conc}

We have investigated the effects of isocurvature perturbations on the 21cm emission from MHs at high redshifts. Our results showed that if the isocurvature power spectrum is flat ($n_s^{\rm iso}\approx1$), the 21cm MH signal (as measured by the \textit{rms} differential brightness temperature)  changes only by less than a few percent around its peak. However, strongly blue-tilted spectrum ($n_s^{\rm iso}\approx3$) gives rise to a significant increase in the amplitude of the 21cm signal compared with the adiabatic case. The next generation of large radio telescopes such as the SKA and FFTT has the potential to detect these 21cm imprints from a blue isocurvature spectrum.

The characteristic signatures of isocurvature perturbations on the MH abundances were explored in detail. In particular, we found an unexpected deficit in small-mass MHs when the isocurvature fraction increases beyond a certain threshold. We explained this phenomenon in terms of the incorporation of small-mass MHs into larger haloes. 

A detailed Fisher-matrix analysis was performed to study quantitatively how  the 21cm signals from MHs can constrain the isocurvature amplitude and spectral index. We found that if the isocurvature spectrum is flat, 1) the combination of  CMB and 21cm experiments fares no better than the CMB alone, 2) the CDM and baryon types of isocurvature fluctuations are unlikely to be distinguishable, even with the futuristic CMBPol+FFTT specifications. However, if $n_s^{\rm iso}\approx3$, there are realistic prospects for distinguishing between different isocurvature types, but only if the 21cm signal from redshifts up to $\sim40$ can be utilised. Some physical models which predict very blue isocurvature spectrum with $\nsiso=2-4$ are discussed in \cite{Kasuya:2009}.

Two sources of uncertainty in the MH population were discussed, namely, the halo mass function, and the mass range of MHs. The amplitudes of the 21cm emission from MHs were shown to be fairly sensitive to the halo mass function, although signals from a blue spectrum remain strong enough to be detected by the SKA and FFTT regardless of the mass function. We also explored the uncertainty in the minimum MH mass, and showed that increasing $M_{\rm min}$ suppresses the 21cm signal over a large range of redshifts, especially at high redshifts where an order-of-magnitude suppression was seen.

For the two sources of uncertainty above, we also obtained the error contours in the  $r_{\rm cdm}-r_{\rm bar}$ and the $r_{\rm cdm}-n_s^{\rm iso}$ planes. These constraints are sensitive to the choice of the mass function: The Press-Schechter prescription, in particular, can be construed as giving overly optimistic constraints.  Increasing  $M_{\rm min}$ suppresses the MH signal strongly at high redshifts, hence the error contours are also significantly widened. 
\\

In summary, the fluctuations of the 21cm emission from MHs are a viable tool in the search for isocurvature perturbations, and have the potential to rule out inflation models which predict a very blue-tilted isocurvature spectrum. When combined with CMB constraints, future 21cm experiments have the potential to distinguish between the CDM and baryon types of isocurvature perturbations. This will be extremely useful in the understanding of physics in the inflationary era.

Our analysis focused on uncorrelated CDM and baryon isocurvature modes, but it is plausible that there may be a nontrivial correlation between the two. Such a correlation gives rise to additional degrees of freedom. In future work, it will be interesting to explore the parameter space allowed by certain inflationary theories which predict correlated isocurvature modes.

\bigskip

\noindent [Note: prior to the publication of this work, we became aware of the work by \cite{Sekiguchi:2013}, which significantly overlaps with our work. The conclusions in their work are similar to ours.]


\section*{Acknowledgements}
We would like to thank J. Silk and H. Tashiro for useful discussions and
comments. We acknowledge support from JSPS (Japan Society for
Promotion of Science) Fellows (YT); 
This work is supported in
part by a Grant-in-Aid for Nagoya University Global COE Program ``Quest
for Fundamental Principles in the Universe: from Particles to the
Solar System and the Cosmos'', from the Ministry of Education, Cluster,
Sports, Science, and Technology (MEXT); Kobayashi-Maskawa Institute
for the Origin of Particles and the Universe; Nagoya University for
providing computing resources useful in conducting the research
reported in this paper.

\bibliography{isomh}

\appendix

\bsp

\label{lastpage}

\end{document}